\documentstyle[prl,aps,psfig]{revtex}
\def\vecr{{\rm\bf r}}
\def\Imag{{\rm Im}}
\def\trace{{\rm Tr}}
\def\ioverh{{\frac{i}{\hbar}}}

\begin{document}
\bibliographystyle{prsty}
\title{Semiclassical Theory of Integrable and Rough Andreev Billiards}
\author{W. Ihra$^1$, M. Leadbeater$^{1,2}$, J. L. Vega$^1$
         and K. Richter$^1$ \medskip}
\address{$^1$Max-Planck-Institut f\"ur Physik komplexer Systeme, 
         N\"othnitzer Stra{\ss}e 38, 01187 Dresden, Germany, \\
          $^2$Dipartimento di F\'{\i}sica, Universit\'a di Roma III,
           Via della Vasca Navale 84, 00146 Roma, Italy}
\medskip
\maketitle

\begin{abstract}
We study the effect on the density of states in mesoscopic ballistic
billiards to which a superconducting lead is attached. 
The expression for the density of states is 
derived in the semiclassical S-matrix formalism shedding insight into the
origin of the differences between the semiclassical theory
and the corresponding result derived from random matrix models.
Applications to a square billiard geometry and billiards with 
boundary roughness are discussed. The saturation of the quasiparticle
excitation spectrum is related to the classical
dynamics of the billiard. The influence of weak magnetic
fields on the proximity effect in rough Andreev billiards is discussed
and an analytical formula is derived. The semiclassical theory
provides an interpretation for the suppression of the proximity
effect in the presence of magnetic fields as a coherence effect
of time reversed trajectories, similar to the weak localisation 
correction of the magneto-resistance in chaotic mesoscopic systems.
The semiclassical theory is shown to be in good agreement with
quantum mechanical calculations. \\

\parindent=0pt
PACS numbers: 05.45+b, 74.50+r, 74.80Fp
\end{abstract}

\bigskip
\bigskip
\leftline{\bf 1 Introduction}
\bigskip

A superconductor in proximity to a normal conductor affects the
spectral density of quasiparticle excitations in the
conductor. Recent technological advances in building 
very clean conductors of mesoscopic size have led to consider
this proximity effect not only in the dirty disorder limit,
where it is known to play an important role in many transport properties
\cite{Mil68,Vol94,Bel96,Lamb98}, but also
in ballistic mesoscopic samples \cite{Mel96,Mel97,Lod98,Schom99}
in proximity to a bulk superconductor. These systems are often
modelled by ballistic billiards. Billiard shaped structures
connected to a superconductor have been coined Andreev billiards
\cite{Kosz95}. It was shown in
\cite{Mel96,Mel97} that the form of the quasiparticle excitation
spectrum in the Andreev billiard depends crucially on whether
its classical dynamics is integrable or chaotic. 
A semiclassical interpretation of these results 
based on the Eilenberger Green's function \cite{Eil68} was given in
\cite{Lod98}. The semiclassical approach has also been extended to
mesoscopic samples with a mixed classical phase space \cite{Schom99}. 

The aim of this contribution is twofold: We first present a derivation of
the semiclassical result for the quasiparticle excitation spectrum 
based on the semiclassical scattering matrix approach as
pioneered by Smilansky and co-workers\cite{Smi92,Smi94} which
allows for a transparent physical interpretation of the result of
\cite{Lod98} in terms of multiple Andreev scattering events \cite{Andr64}.
Secondly, we discuss a number of features of the quasiparticle excitation
spectrum in an integrable square billiard and in a square billiard
with surface roughness. These features include
the saturation of the quasiparticle excitation spectrum in the square
billiard and a semiclassical explanation of 
the effect a weak magnetic field has on the quasiparticle
excitation spectrum in billiards with surface roughness.

The semiclassical expression for the quasiparticle density of states
will be derived in Section 2. The strength of this approach lies
in the fact that it allows to
see on which level approximations enter semiclassically. The
semiclassical theory of chaotic Andreev billiards
predicts an exponential suppression of the density
of states near the Fermi energy \cite {Lod98,Schom99}. In contrast
a random matrix modelling \cite{Mel96,Mel97} leads to the result
of a gap in the density of states in an energy
interval above the Fermi energy. The scattering matrix approach
elucidates one possible origin of the differing results.

In Sec. 3.1 we discuss the quasiparticle excitation spectrum of a
square Andreev billiard in detail. While previous papers concentrated
on the linear rise of the spectrum above the Fermi energy 
\cite{Mel96,Mel97,Lod98} we focus on the 
saturation of the excitation spectrum at higher energies. We show that
the saturation can be related to the probability distribution of short
classical paths hitting the superconducting parts of the billiard
boundaries. 
In Sec. 3.2 we discuss results for square billiards with additional surface
roughness. Surface roughness is modelled as isotropic scattering of
electrons and holes from the normal parts of the billiard boundary.
An exponential suppression of the density of states near the Fermi
energy is observed in complete analogy to chaotic billiards \cite{Schom99}.
Finally in Sec. 3.3 the effect of an additional weak magnetic field
on the quasiparticle excitation spectrum in the rough billiard is considered.
We predict an enhanced density of states
compared to the field free case for energies near $E_F$
and give analytical semiclassical
expressions. The semiclassical theory provides a clear interpretation
of this result as an effect of destructive interference between
reversed paths in the presence of a magnetic field. Again the
semiclassical predictions are in good agreement with quantum
mechanical calculations.
\bigskip

\leftline{\bf 2 Theory}
\bigskip

\begin{figure}
\begin{minipage}{16.6cm} 
  \centerline{\large Fig. 1}
  \vspace{1.0cm}
 \centerline{\psfig{figure=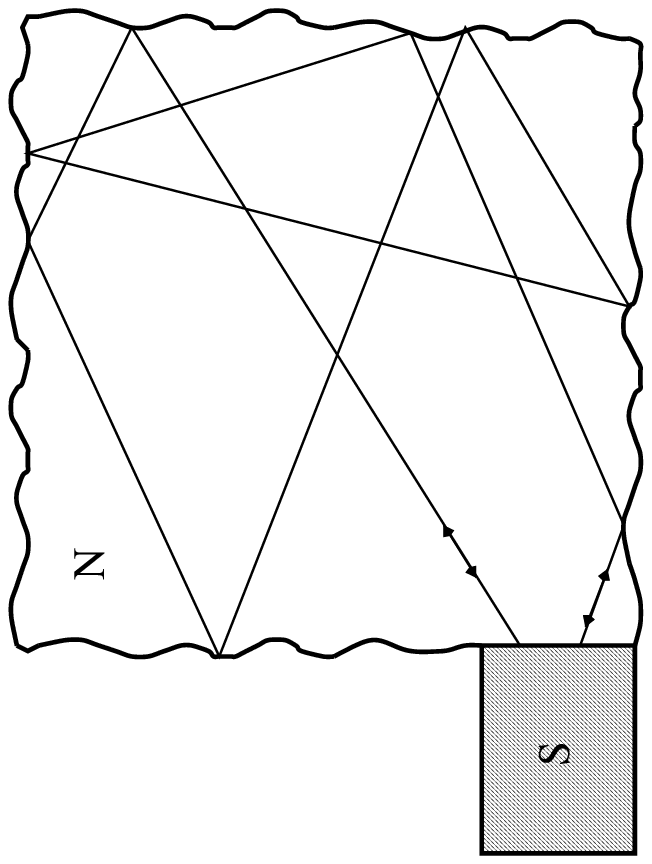,angle=-90,width=8.0cm}}
  \vspace{1.0cm}
  \caption[Short title]{Square billiard with rough
                        boundaries. Roughness is modelled by isotropic
                        scattering when an electron or hole hits the 
                        normal billiard boundary (reflection in any
                        direction with equal probability) in contrast 
                        to specular reflection at a smooth boundary.
                        At the SN boundary electrons are 
                        retro-reflected (Andreev reflected)
                        as holes and vice versa.}
   \label{fig1}
\end{minipage}
\end{figure}

In a quasi-classical picture the 
effect of a superconducting lead coupled to a mesoscopic billiard
manifests itself in the process known as Andreev reflection \cite{Andr64}:
At the superconducting parts of the billiard boundary electron-like
quasi-particles are retro-reflected with opposite velocities
as holes and vice versa (see Fig. \ref{fig1}, and Refs. \cite{Lamb98,Been94}
for a detailed description). In contrast 
electrons and holes are specularly reflected at those parts
of the billiard boundary which are not in contact to the
superconductor if the billiard boundary is smooth. For a billiard
with surface roughness we will take the possibility of isotropic
scattering off the normal walls of the billiard into account. This
is also indicated in Fig. \ref{fig1}.

The local quasiparticle density of states (DOS) $d(E,{\bf r})$ 
of an Andreev billiard is defined as the density of states at 
positive energy $E$ above the Fermi energy $E_F=0$, weighted with the
corresponding electron-like component of the wave
function at point ${\bf r}$. It is given by\cite{Lod98}
\begin{equation}\label{eq:locdens}
     d(E,{\bf r}) =  \frac{d_N}{A} \int_0^{\pi} d\phi 
     \sum_n \delta\left(\frac{E
     L(\phi)}{\hbar v_F} -(n+\frac{1}{2})\pi\right) \,.
\end{equation}
$A$ denotes the area of the billiard and $v_F$ denotes
the Fermi velocity. It is assumed that 
$E$ is much smaller than the pair
potential $\Delta(\vecr)$ in the superconducting lead and
that the lead supports a large number $N\gg 1$ of
classically allowed transverse channels. In the billiard the pair
potential $\Delta({\bf r})$, which couples electron and hole like
states, vanishes identically. Due to the condition
$E\ll\Delta$ electrons in the billiard which hit the superconducting
lead get reflected as holes thus quantum mechanically forming
electron-hole quasiparticle states.
$L(\phi)$ is the length of the trajectory which passes the point
${\bf r}$ in the billiard under an angle $\phi$ between two successive
bounces with the superconducting boundary and $d_N = m A/(2\pi\hbar^2)$ 
is the average density of states in the isolated billiard. We further 
assume a perfectly
transmitting boundary between the billiard and the lead if the lead
is in the normal state (no potential difference), which has the effect
that the probability for Andreev reflection equals one if the lead is
in the superconducting state. (A situation including the probability
for Andreev reflection as well as normal reflection between SN boundaries was
taken into account in a somewhat different context and geometry
e.g. in\cite{Wees92,Morp97}). 

The total quasiparticle DOS in the billiard is obtained by
integrating (\ref{eq:locdens}) over the area of the billiard system.
The integral over the billiard
area can be converted into an integral over initial starting angles
$\alpha$ and positions $y$ of trajectories along the lead of width $w$.
The resulting expression is
\begin{equation}\label{eq:dosgen}
            d(E) = \frac{d_N}{A} \int \limits_{0}^{w} dy
               \int\limits_{-1}^{1} d(\sin\alpha)
            \int\limits_{0}^{L(y,\alpha)} ds \sum_{n=0}^{\infty}
              \delta\left(\frac{E
            L(y,\alpha)}{\hbar v_F} -(n+\frac{1}{2})\pi\right) \,,
\end{equation}
where $s$ is the local variable measuring the
length along a trajectory. If one additionally assumes an ergodic
distribution of trajectories in the initial conditions
$y$ and $\sin\alpha$ on the boundary the expression
for the density of states can be rewritten in terms of the 
probability distribution $P(L)$ of trajectories of length $L$
between two successive bounces with the SN boundary. It is then
\cite{Lod98,Schom99}
\begin{equation}\label{eq:dosrough}
   d(E) = \frac{2 d_N w}{A} \int_{0}^{\infty} dL\, P(L)\, L 
          \sum_n \delta\left(\frac{E L}{\hbar v_F} -
          (n+\frac{1}{2})\pi\right) \,.
\end{equation}
The assumption of an ergodic distribution is justified in the case
of rough billiards when the time $t_{\rm erg}$ on which the classical
motion in the billiard becomes ergodic is much smaller than the
mean escape time $\tau_{\rm esc}$ of a particle from the billiard into
the SN-lead. It is also a good approximation for the integrable
square billiard with a large number of open channels $N\gg 1$
because then the distribution of initial conditions in $\sin\theta$
is quasi-continuous. For a small number of channels 
it is more appropriate to resort to the continued
fraction evaluation of (\ref{eq:dosgen}) described in Sec. IIIa,
which takes the quantization of angles explicitely into account.

We now aim to derive (\ref{eq:dosrough}) from the semiclassical
scattering matrix approach for quantisation\cite{Smi94} and to
understand the approximations which lead to (\ref{eq:dosrough})
on the semiclassical level. The scattering matrix approach for
quantisation of a billiard system which is opened via a lead
starts with a trace formula for the density of states $d(E)$ 
which involves two contributions: The first term $d_R(E)$ is the
resonance density of states in the corresponding billiard opened
via the lead. 
The second contribution takes the coupling to the superconductor into account
and is expressed as a sum over traces of powers of the
scattering matrix which relates incoming and outgoing transverse modes
in the lead. The result is:
\begin{equation}\label{eq:scattdos}
  d(E) = d_R(E) - \frac{1}{\pi} \lim_{\eta\to 0} \Imag
        \frac{\partial}{\partial E} \ln\det\left[1- S_A S_N(E+i\eta)\right] \,
\end{equation}
where the total scattering matrix for the Andreev billiard
is composed of a product of
a 2N times 2N normal scattering matrix $S_N(E)$ and an Andreev scattering
matrix $S_A$ of equal dimension. The two scattering matrices are given
as\cite{Fra96} 
\begin{equation}\label{eq:Smatrix}
 S_N(E) = \left(\begin{array}{cc}
             S(E) & 0     \\        
              0   & S^{*}(-E)
            \end{array}\right)   \quad {\rm and} \quad\quad
 S_A = \left(\begin{array}{cc}
             0 & -i     \\        
            -i  & 0
            \end{array}\right).
\end{equation}
The normal scattering matrix $S_N$ has block diagonal
structure where the two N times N blocks describe electron and hole
scattering between the channel modes. The Andreev scattering matrix
$S_A$ couples electrons and holes at the SN-interface. The
sub-diagonal elements are N times N unit matrices with an additional phase
of $(-i)$. We neglect the weak energy dependence of the Andreev
scattering matrix which is valid in the deep sub-gap regime $E\ll\Delta$
\cite{Been94,Been91}. In terms of the electron and hole scattering
matrix the density of states takes the form
\begin{equation}\label{eq:traceform}
  d(E) = d_R(E) - \frac{1}{\pi} \Imag\sum_{m=1}^{\infty} 
        \frac{(-1)^m}{m} \trace
        \frac{\partial}{\partial E} \left[S(E) S^{*}(-E)\right]^m \,,
\end{equation}
where (\ref{eq:scattdos}) has additionally been expanded into a sum over
traces of powers of the scattering matrix. An averaged density can be 
obtained by different procedures:
One can average $d(E)$ over a classically small
interval of Fermi energies $E_F$ or one
can average $d(E)$ over different realizations of the rough billiard
(with fixed area $A$). In any
case averaging of the resonance density gives the average density
$d_N$ of the isolated billiard, and the average quasiparticle density
is
\begin{equation}\label{eq:avdens}
  d_{av}(E) = d_N(E) - \frac{1}{\pi} \Imag\sum_{m=1}^{\infty} 
        \frac{(-1)^m}{m} \trace
        \frac{\partial}{\partial E} \langle\left[ S(E)
                S^{*}(-E)\right]^m\rangle\,.
\end{equation}
Averaging is denoted by brackets $\langle\cdots\rangle$.
Equation (\ref{eq:avdens}) is the
starting point for the derivation of expression
(\ref{eq:dosrough}) within the semiclassical scattering matrix approach.

Semiclassically the elements $S_{nn'}(E)$ of the electron scattering
matrix are expressed as a sum over classical orbits. Each orbit
contributes with an amplitude $A_j$ and a phase $S_j$. The
resulting expression is\cite{Smi92,Mil74}
\begin{equation}\label{eq:semsmat}
    S_{nn'}(E) = \sum_j A_j(n\to n')
         \exp\left[\frac{i}{\hbar}S_j(E,n\to n')-i\frac{\pi}{2}\nu_j\right]\,,
\end{equation}
where $\nu_j$ is an additional integer Maslov index.
The amplitude pre-factor is given explicitely by
\begin{equation}\label{eq:amplprefact}
   A_j(n\to n') = \left(\frac{\hbar}{2\pi}\right)^{1/2} 
      \left\vert\frac{\partial I'(E)}{\partial\theta} \right\vert^{-1/2}
\end{equation}
where $I'=\hbar n'$ is the action of the final transverse motion in
the lead. Only those paths contribute which enter the billiard at the
SN-boundary with fixed quantised angle $\pm\sin\theta=n\pi/(k_F W)$ and
return to the boundary with angle $\pm\sin\theta'=n'\pi/(k_F W)$. In
terms of the initial position $y$ along the SN boundary and the angle
$\theta'$ with which the trajectory returns the amplitude can be
written as
\begin{equation}\label{eq:amplprefact2}
   A_j(n\to n') = \frac{1}{w}\sqrt{\frac{\pi}{2k_F}}
      \left\vert\frac{\partial y}{\partial(\sin\theta')}
\right\vert^{1/2} \,.
\end{equation}

To proceed further we resort to the physical picture of Andreev
reflection of electrons into holes and vice versa at the SN-boundary.
We first observe that the traces contain products of alternating
electron scattering matrices $S(E)$ and hole scattering matrices
$S^{*}(-E)$. The energy $E$ above the Fermi level is large with
respect to the mean level density $\delta\equiv d_N^{-1}$ of the isolated
billiard, but is classically small. Similar to the semiclassical
evaluation of density-density
correlator\cite{Ber85,Ozo88,Arg93} we expand the phase
around the Fermi energy as $S_j(\pm E) \simeq S_j(0) \pm E T_j(0)$
where $T_j$ is the return time of the orbit to the SN-interface.
Additionally the amplitudes are only slowly varying functions of the energy
and are evaluated at the Fermi energy.

We demonstrate how to evaluate the sum over traces of products of
the scattering matrices for the $n=1$ term. Generalisation to higher
order terms is straightforward. Matrix elements of products of an 
electron and a hole scattering matrix have the form
\begin{equation}\label{eq:elholeprod}
  [S(E)S^{*}(-E)]_{nn'} = \sum_{n''} \sum_{j,k} A_j(n\to n'')
        A_k^{*}(n''\to n') \exp\left[\ioverh(S_j-S_k)\right]
         \exp\left[\ioverh E(T_j+T_k)\right] \,.
\end{equation} 
Upon averaging $[S(E)S^{*}(-E)]_{nn'}$ over the Fermi energy or
different realizations of boundary roughness in the case of rough
billiards only diagonal terms $j=k$ contribute to the sum. The
diagonal approximation is further justified by the physical picture
of Andreev reflection, which means that the reflected hole orbit
retraces the electron orbit and vice versa. However it must be
emphasised that Andreev reflection is exactly fulfilled only at
the Fermi energy $E_F=0$, and in an exact treatment deviations
from Andreev reflection at finite energies have be taken into 
account.

The diagonal approximation implicates that
the initial and final channel indices are equal, $n=n'$, and the
product matrix (\ref{eq:elholeprod}) is diagonal:
\begin{equation}\label{eq:snn}
 [S(E)S^{*}(-E)]_{nn'} = \delta_{nn'} 
                           \sum_{n''}\sum_j \left\vert A_j(n\to n'')
                         \right\vert^2 \exp\left(2\ioverh E T_j\right)
                          \,.
\end{equation}
In the semiclassical limit the summation over intermediate channels
$n''$ can be transformed into an integral over angles:
$\sum_{n''} k_F w/\pi \int_{-1}^{1} d(\sin\theta')$.
Using the expression (\ref{eq:amplprefact2}) for the amplitudes 
one arrives at the final expression
\begin{equation}\label{eq:eqholeprod2}
 [S(E)S^{*}(-E)]_{nn'} =  \delta_{nn'} \frac{1}{2w} \int_{0}^{w} dy
                         \exp(2\ioverh E T(y)) \,.
\end{equation}
Taking the trace amounts to another integration over the angle in the
semiclassical approximation and one has
\begin{equation}\label{eq:eqholeprod3}     
 \trace[S(E)S^{*}(-E)] =  \frac{k_F}{2\pi} \int_{-1}^{1}
                d(\sin\theta) 
                 \int_{0}^{w} dy
                \exp(2\ioverh E T(y)) \,.
\end{equation}
Within the diagonal approximation traces of higher powers of products
of electron and hole scattering matrices are easily shown to be
\begin{equation}\label{eq:eqholeprod4}     
 \trace[S(E)S^{*}(-E)]^m =  \frac{k_F}{2\pi} \int_{-1}^{1}
                d(\sin\theta)
                 \int_{0}^{w} dy
                \exp(2\ioverh m E T(y)) \,.
\end{equation}
Using the trace formula (\ref{eq:avdens}) this gives
\begin{equation}\label{eq:averagedos}
    d_{av}(E) =  d_N + \frac{k_F}{2\pi^2\hbar} \sum_{m=1}^{\infty}
          (-1)^m \int_{0}^{w} dy \int_{-1}^{1} d(\sin\theta) T 
            \cos(\frac{2mT}{\hbar} E)  \,.
\end{equation}
If now Poissonian summation is used the equivalence
of (\ref{eq:averagedos}) and the Bohr-Sommerfeld like
expression (\ref{eq:dosgen}) for the average density of states in the
Andreev billiard, which was derived on the basis of the Eilenberger
equation for the Green's function \cite{Mel96}, can easily be seen.

The scattering matrix approach gives a clear and intuitive
interpretation of how the coupling of the billiard to superconducting
leads modifies the average density of states: It is given as a sum of the
average density of states of the isolated billiard (the Weyl term)
plus a sum of multiple Andreev reflections of electron into hole
states. Off-diagonal corrections arise due to the fact that the condition of
Andreev reflection is exactly fulfilled only at the Fermi energy.
The presence of non-diagonal contributions points towards a possible
explanation
for the difference between the random matrix result for a chaotic
Andreev billiard \cite{Mel96,Mel97}, which predicts a gap in the
quasiparticle excitation spectrum for billiards with a chaotic
classical phase space, and the semiclassical theory which leads to
an exponential suppression \cite{Lod98,Schom99}. 
Pairs of non-identical trajectories $k\neq
j$ which follow the same path along a segment in real space before
they depart due to slightly different initial conditions and the
chaotic dynamics may have similar actions and therefore also survive
the averaging of products of electron and hole scattering matrices,
Eq. (\ref{eq:elholeprod}) and powers of it \cite{Frah}. An extended
diagonal approximation which takes the contribution of paths into
account which are not exactly related by time-reversal symmetry has recently been
proposed in the context of weak localisation \cite{Alei96,Smi98}.
The influence of off-diagonal contributions arising in the
semiclassical approach to the density of states in Andreev billiards
remains a topic to be investigated. 

Finally it should be emphasised that averaging has been performed
on classically small scales. Fluctuations of 
$d_{av}(E)$ can still appear on classical energy scales due to
fluctuations in the length distribution probability $P(L)$. In the
following we will skip the index. $d(E)$ then also denotes the
density of states after the above described averaging.
\bigskip

\leftline{\bf 3 Results}
\bigskip

In the following we present results of quantum mechanical calculations
for the average density of states in Andreev billiards and compare
them with the semiclassical theory. We will focus on the difference
between results for a square billiard which is integrable and a square
billiard with rough boundaries. As an effect of rough boundaries 
electron and hole trajectories can be scattered into arbitrary
directions when
they hit the normal boundary. We will then consider the effect of a
weak magnetic field on the density of states and derive a
semi-analytical expression for the magnetic flux and energy dependence
of the density of states.

To numerically model the structure of Fig.\ref{fig1} we consider 
a ballistic normal region
connected to a clean superconductor.
For our numerical calculation we use 
a tight-binding version of the Bogoliubov-de Gennes Hamiltonian: 
\begin{equation}
\pmatrix{ H_0  & \Delta \cr \Delta^*  & 
-H_0 }\pmatrix{u \cr v }\, =\, 
E\pmatrix{u \cr v }.
\label{bdgeq}
\end{equation}
In this equation 
$H_0=\sum_{i}|i\rangle \epsilon _{i}\langle i|-t\sum_{\langle 
ij\rangle }|i\rangle \langle j|$ is the standard single--particle 
Anderson model
with $\langle ij\rangle$ denoting pairs of nearest neighbour sites
and $\Delta =\sum_{i}|i\rangle 
\Delta_{i}\langle i|$ is the superconducting order parameter.
The billiard has width $M$ and length $L$ 
(in units of the lattice constant $a$). The coupling to the superconductor is of
width $w$. 
With the exception of the billiard boundary, the diagonal
matrix elements $\epsilon _{i}=\epsilon _{0}$, with $\epsilon_{0}$ 
chosen to ensure that the Fermi level is away from the van Hove
singularity in the band centre ($\epsilon_0=0$). To model the boundary 
roughness, for sites on the boundary we randomly
choose $\epsilon_i=10^4$ or $\epsilon_i=\epsilon_0$ with equal probability.
In the absence of a magnetic field,
the off-diagonal matrix elements $t$, which
determine the width of the energy band (the band-width is $8t$), 
are set to $1$ throughout the system. The effects of a magnetic field are modelled
by including a Peierls' phase factor into these off-diagonal elements 
in the billiard region. To compute
the density of states, a numerical decimation
technique is employed \cite{decim}. This has been used extensively over recent
years to discuss transport and density of states properties of hybrid
systems\cite{ml}.
\bigskip

\leftline{\bf 3.1 Square billiards}
\bigskip

The asymptotic length distribution of trajectories in square billiards
typically follows a power law of the form $P(L) \sim L^{-3}$. This
result was used in \cite{Mel96} to show that the density of
quasiparticle states is a linear function of the energy of
quasiparticle excitations for $E\ll\Delta$. The slope of the linear
behaviour is related to the proportionality constant of the asymptotic
length distribution. The above distribution however leads to a linear
growth of the density of states at all energies and cannot reproduce
the large energy limit $d(E)\to d_N$ for the quasiparticle density.
We show that the crossover to $d_N$ is related to a modified
length distribution $P(L)$ for small lengths. For widths $w$ 
of the superconducting channel much smaller than the length $a$
of the billiard the length distribution reaches approximately a
plateau for small lengths. We introduce two different length
distribution functions and compare numerical results to the analytical
results derived from the smooth length distribution functions. 

It is useful to express the density of states in terms of scaled
lengths and energies. Introducing the lengths $L_T = \pi A/w$  and
$E_T=\hbar v_F/(2 L_T)$ which are sometimes referred to as the
Thouless length and the Thouless energy \cite{Mel96,Mel97,Lod98}, 
lengths $l=L/L_T$ and
energies $\epsilon=E/E_T$ are expressed in terms of these units.
The meaning of $L_T$ and $E_T$ will become clear for chaotic
billiards (Sec. IIIB), where $L_T$ is the average length of trajectories 
before they escape from the billiard. We use the same units of
length and energy for the integrable square billiard in order to
compare results with the rough billiard later.
Expressed in scaled quantities the density of states has the form
\begin{equation}\label{eq:scaldens}
 \frac{d(\epsilon)}{d_N} = \pi\int_{0}^{\infty}
     dl P(l) l \sum_{n=0}^{\infty} 
      \delta\left(\frac{\epsilon l}{2} - (n+\frac{1}{2})\pi\right) 
\end{equation}
or equivalently 
\begin{equation}\label{eq:scaldens2}
  \frac{d(\epsilon)}{d_N} = \frac{(2\pi)^2}{\epsilon^2}
      \sum_{n=0}^{\infty} (n+\frac{1}{2})  P(l_n) \,.
\end{equation}
We introduce two different length distribution functions. The first
length distribution function $P_c(L)$ approximates the probability 
by a constant for lengths $l<l_c$:
\begin{equation}\label{eq:Plc}
        P_c(l) =
           \left\{ \begin{array}{ll} 
               C_c/l_c^3    & \quad\quad  l < l_c \\
               C_c/l^3      & \quad\quad
            l \geq  l_c
          \end{array} \right.
\end{equation}
The second distribution function has a smooth crossover to a constant
probability at small $l$:
\begin{equation}\label{eq:Pls}
        P_s(l) =
                \frac{C_s}{l^3 + l_s^3}   \,.
\end{equation}

We first discuss the length distribution function $P_c(l)$.
For a given energy $\epsilon$ there is a value $n_0$ in the sum
(\ref{eq:scaldens2}) given
by $n_0 = \epsilon l_c/(2\pi)-1/2$ so that for $n>n_0$ the
algebraic tail of the length distribution for $P_c(l_n)$ applies
while for $n<n_0$ $P_c(l_n)$ is constant. The density of states
is 
\begin{equation}\label{eq:scaldens3}
   \frac{d(\epsilon)}{d_N} = C_c \frac{(2\pi)^2}{\epsilon^2}
       \left[ \frac{1}{l_c^3} \sum_{n=0}^{n_0-1} (n+\frac{1}{2})
            + \frac{\epsilon^3}{(2\pi)^3} \sum_{n=n_0}^\infty
           \frac{1}{(n+\frac{1}{2})^2} \right] 
\end{equation}
which after summation and using the relation between $n_0$ and
$\epsilon$ becomes 
\begin{equation}\label{eq:scaldens4}
    \frac{d(\epsilon)}{d_N} = C_c \left[ \frac{2\pi^2}{l_c^3\epsilon^2}
            \left(\frac{\epsilon l_c}{2\pi}-\frac{1}{2}\right)^2
             + \frac{\epsilon}{2\pi} \psi'
               \left(\frac{\epsilon l_c}{2\pi}\right) \right]
             \quad {\rm for} \quad \epsilon > \pi  \,.
\end{equation}
$\psi'$ is the derivative of the Digamma function.
This expression for the spectral density is valid for $\epsilon >
\pi$, otherwise $n_0=0$ and the first part in the sum does not
contribute. For $\epsilon < \pi$ the result is
\begin{equation}\label{eq:scaldens5}
    \frac{d(\epsilon)}{d_N} = C_c \frac{\pi}{4} \epsilon \,, 
\end{equation}
which shows the known linear behaviour for small energies in square billiards
\cite{Mel96}. Using $\psi'(x) \to 1/x$ as $x \to
\infty$ the asymptotic behaviour for large energies is given by
\begin{equation}\label{eq:scaldens6}
    \frac{d(\epsilon)}{d_N} \to \frac{3C_c}{2l_c} \quad{\rm as} \quad
          \epsilon\to\infty  \,.
\end{equation}
The values of the free parameters are determined by the requirements
that (a) the probability distribution must be normalised:
$\int_{0}^{\infty} P(l) dl = 1$ and (b) for large excitation energies
the influence of the coupling to the superconductor vanishes and thus 
$d(\epsilon)\to d_N$ as $\epsilon\to\infty$. This gives the values $l_c=1$ and
$C_c=2/3$. It is seen that these requirements automatically determine
the proportionality constant $C_c$ of the asymptotic power law tail
of the length distribution (\ref{eq:Plc}) and thus the slope
$C_c \pi/4$ of the linear behaviour (\ref{eq:scaldens5})
of the low energy spectral
density. The value $\pi/6$ is somewhat smaller than the value $2/\pi$
given by the authors of \cite{Mel96} who based their calculations on 
a numerically observed value of $C=8/\pi^2$ but did not take a
crossover of $P(l)$ to a flat distribution at some value $l_c$ into
account. Note that one
could adjust $C$ independently to a given value if one introduced
a three parameter function for the length distribution $P(l)$, i.e.
by giving a cut-off length $l_{*}$ below which $P(l_{*})=0$. Such a
cut-off length parameter would quite naturally be of order $l_{*}
\approx 2a$. We will however not deal with this alternative here as when
using a three parameter length distribution probability formulas
become rather involved.

The length distribution (\ref{eq:Plc}) is easy to handle and gives
results which agree well with quantum mechanical calculations. It leads
however to an unphysical discontinuity of the derivative of
$d(\epsilon)$ at $\epsilon=\pi$. The second length distribution
(\ref{eq:Pls}) we used is free of this feature. Proceeding in the
same way as outlined above the spectral density of quasiparticle
states is given by
\begin{equation}\label{eq:scaldens7}
    \frac{d(\epsilon)}{d_N} = \frac{C_s}{2\pi} \epsilon
     \sum_{n=0}^{\infty} \frac{n+\frac{1}{2}}{
        \left(\frac{\epsilon l_s}{2\pi}\right)^3 + 
             \left(n+\frac{1}{2}\right)^3} \,.
\end{equation}
The requirement of normalisation of $P_s(l)$ leads to $C_s = 3\sqrt{3}
l_0^2/(2\pi)$.
Further evaluation is only possible in the small and large energy limits.
The small energy limit ($\epsilon l_s \ll 2\pi$) gives again the
result (\ref{eq:scaldens5}) with $C_c$ replaced by $C_s$. In the
opposite high energy limit the
summation can be replaced by integration over the continuous variable
$x=2\pi/(\epsilon l_s) (n+1/2)$, leading to
\begin{equation}\label{eq:scaldens8}
    \frac{d(\epsilon)}{d_N} =
     \frac{C_s}{l_s} \int_{x_0}^{\infty} \frac{x}{1+x^3} dx \,,
            \quad\quad x_0 = \frac{\pi}{\epsilon l_s} \,,
\end{equation}
and finally
\begin{equation}\label{eq:scaldens9}
 \frac{d(\epsilon)}{d_N} =
            \frac{C_s}{\pi} \epsilon \,\,
       _2F_1\left(\frac{1}{3},1,\frac{4}{3},-\frac{1}{x_0^3}\right)
               \to \frac{2\pi C_s}{3\sqrt{3}l_0} \quad {\rm as} \quad 
                 \epsilon\to\infty
                \,.
\end{equation}
The values of the parameters of the length distribution function are
given by $l_s=1$ and $C_s=3\sqrt{3}/(2\pi)$. This value of $C$ is very
close to $C=8/\pi^2$ of \cite{Mel96} giving a slope of $3\sqrt{3}/8
\approx 0.6495$ of $d(\epsilon)/d_N$ for small energies
compared to $2/\pi \approx 0.6366$ of \cite{Mel96}. 

\begin{figure}
\begin{minipage}{16.6cm} 
  \centerline{\large Fig. 2}
  \vspace{1.0cm}
  \centerline{\psfig{figure=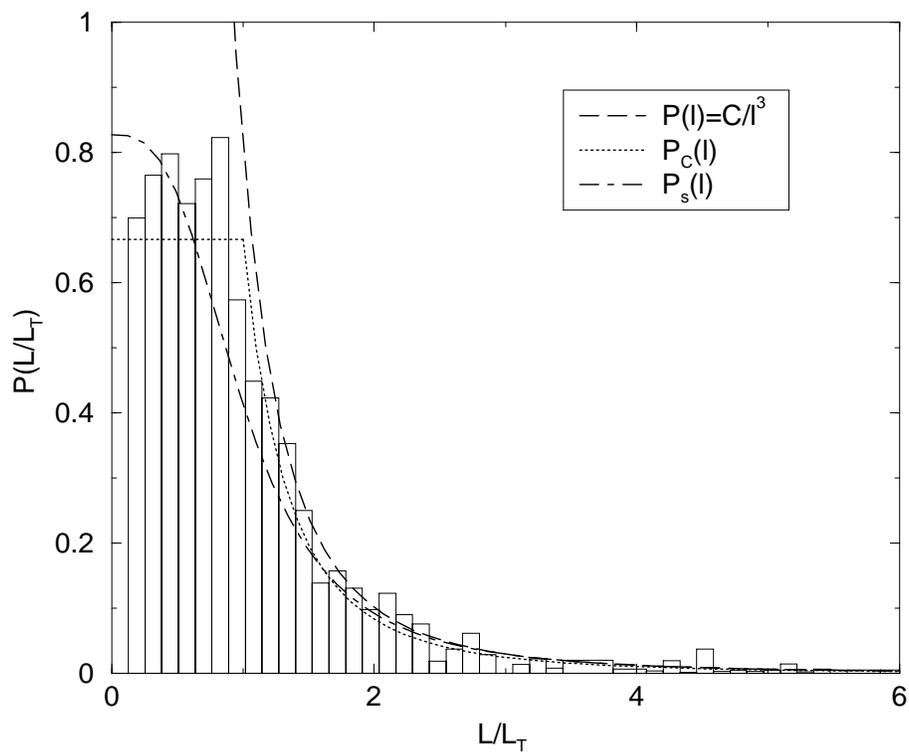,width=12.0cm}}
  \caption[Short title]{Histogram of the length distribution
                        probability $P(L/L_T)$
                        for a square billiard obtained by evaluating
                        Eq. (\ref{eq:doscontfrac}) by the method
                        of continued fractions.
                        Parameters: Length of the billiard sides
                        $a=1$, width of the superconducting lead
                        $w=0.1$. The histogram is an average over 
                        different numbers of channels ranging from
                        $N=96$ to $N=105$. Dashed line: asymptotic
                        power law for large $l$. Dotted line: 
                        $P_c(l)$, Eq.
                        (\ref{eq:Plc}). Dot-dashed line: 
                        $P_s(l)$, Eq. (\ref{eq:Pls}).}
   \label{fig2}
\end{minipage}
\end{figure}

In Fig. \ref{fig2} we present a characteristic example of length
distribution probabilities for a small channel width $w\ll a$. A
plateau in the length distribution function $P(l)$ for $l<1$ is
clearly visible. Here the crossover from the asymptotic
power law for long lengths to the plateau is slightly better
modelled by $P_c(l)$ compared to $P_s(l)$. 

\begin{figure}
\begin{minipage}{16.6cm} 
  \centerline{\large Fig. 3}
  \vspace{1.0cm}
  \centerline{\psfig{figure=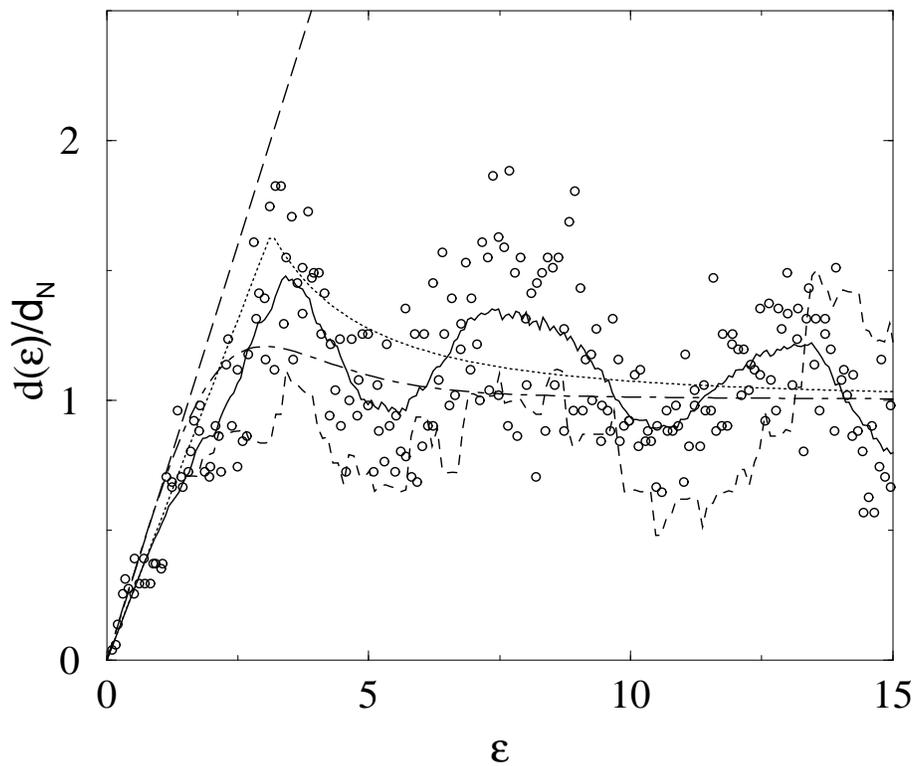,width=12.0cm}}
  \caption[Short title]{Average density of states $d(\epsilon )/d_N$ 
                        as a function of the energy $\epsilon = E/E_T$.
                        The circles are quantum mechanical
                        calculations for a square billiard of side
                        length $a=75$ and channel width
                        $w=25$ ($N=12$). The solid line is
                        a 20-point average over the set of data
                        points. Dashed line: Semiclassical result
                        from Eq. (\ref{eq:doscontfrac}).
                        Dotted line: DOS using $P_c(l)$ from
                        Eq. (\ref{eq:Plc}). Dot-dashed line: DOS using
                        $P_s(l)$ from Eq. (\ref{eq:Pls})
                        Long dashed line: Low energy limit
                        $d(\epsilon)/d_N = (2/\pi)\epsilon$ based on the
                        probability distribution $P(l) =
                        C/l^3$.}
   \label{fig3}
\end{minipage}
\end{figure}

We also calculated the density of states for individual square
billiards starting from the semiclassical formula
(\ref{eq:dosgen}) without employing the assumption of an ergodic
distribution of initial conditions on the SN-boundary. For individual
square billiards this formula is more appropriate than
(\ref{eq:dosrough}) before averaging over Fermi energy
since it does not involve the assumption of an
ergodic distribution of initial conditions of trajectories on the
SN-boundary. In square billiards trajectories with equal length
between two hits with the channel lead are organised in families (for a
discussion see \cite{Pic98}). Orbits with channel index $a=1,\cdots,N$
hit the SN-boundary under an angle $\sin\theta_a = a\pi/(k_F w)$. For
each channel $a$ there is a finite number of orbit families
$\lambda_a$ with different lengths $L_{\lambda_a}$. Each member of
an orbit family carries a weight $\delta_{\lambda_a}$ with which it
contributes to the density of states which is then given by
\begin{equation}\label{eq:doscontfrac}
  d(E) = \frac{d_N}{A} N^{-1} \sum_{a=1}^N\sum_{\{\lambda_a\}}
            \delta_{\lambda_a} L_{\lambda_a} \sum_{n=0}^{\infty}
            \delta\left(\frac{E
            L_{\lambda_a}}{\hbar v_F} -(n+\frac{1}{2})\pi\right)   
\end{equation}
with $\sum_{\{\lambda_a\}} \delta_{\lambda_a} = w$ for the sum of weights of the members
$\lambda_a$ of a single channel $a$. The lengths $L_{\lambda_a}$ 
of contributing trajectory families and their weights
$\delta_{\lambda_a}$ can be very efficiently calculated by means of
an algorithm involving a finite number of continued fractions as
described in \cite{Pic98}. Fig. \ref{fig3} shows the result of a quantum 
mechanical calculation for a billiard with $w/a=1/3$ and $N=12$
channels. The solid line is a 20-point average over the quantum
mechanical data. The dashed line represents the semiclassical result
obtained from Eq. (\ref{eq:doscontfrac}). It follows the general
trend of the quantum mechanical result but is somewhat lower. Also plotted
as dotted and dot-dashed lines are the formulas for the averaged
quasiparticle excitation spectrum obtained from
Eq. (\ref{eq:scaldens4}), (\ref{eq:scaldens5})
and (\ref{eq:scaldens7}). The linear behaviour of the quasiparticle spectrum
for small excitation energies based on the asymptotic length
distribution function $P(l)$ without plateau is plotted as the long 
dashed line. The average of the quantum mechanical result is in good
agreement with the results of Eq. (\ref{eq:scaldens4}),
(\ref{eq:scaldens5}) and (\ref{eq:scaldens7}). Both the
numerical semiclassical result (dashed line) and the quantum mechanical
result show however additionally a pronounced oscillation around the
mean value $d(\epsilon)=d_N$ at energies $\epsilon > 1$. We observed
this trend also for other individual square billiards with different
parameter values of $w/a$ and $N$. The origin of these oscillations
has not yet been identified and deserves further investigations.
\bigskip

\leftline{\bf 3.2 Rough billiards}
\bigskip

In this subsection we apply the semiclassical result (\ref{eq:dosrough})
for the quasiparticle density of states to the square billiard with
additional surface roughness. In the following we discuss the
situation where the width $w$ of the superconducting channel
is much smaller than the length $a$ of one billiard side.
Since a trajectory randomises once it
hits the rough billiard walls due to off-scattering in arbitrary direction
with equal probability the motion becomes ergodic on a time scale much
smaller than the mean time between two hits with the superconducting
part of the boundary. The latter can also be viewed as the escape time
of trajectories from the billiard if it was open along the channel lead. 
The escape probability $P(L)$ of trajectories of length $L$ 
from an open chaotic billiard is known to be given asymptotically
by \cite{Smi92,Bar93} 
\begin{equation}\label{eq:lengthdistrough}
      P(L) = \frac{1}{L_T} \exp\left(-\frac{L}{L_T}\right) 
\end{equation}
with the above defined Thouless length $L_T$ \cite{Mel96}. It is
related to the mean escape time by $\tau_{\rm esc} = L_T/v_F$.
Note that averaging over different realizations of the rough billiard
is effectively taken into account by the introduction of the smooth length
distribution function $P(L)$ by which a probabilistic description on
the classical level is achieved. Numerical calculations confirm the
form (\ref{eq:lengthdistrough}) of $P(L)$ for long
trajectories. Using (\ref{eq:lengthdistrough}) the average density
of states can be calculated from Eq. (\ref{eq:dosrough}) 
by evaluating the delta function integral directly and summing over
$n$. The result is 
\begin{equation}\label{eq:dosforrough}
        d(\epsilon) = d_N \left(\frac{\pi}{\epsilon}\right)^2 
            \frac{\cosh\left(\frac{\pi}{\epsilon}\right)}
                 {\sinh^2\left(\frac{\pi}{\epsilon}\right)}
                  \quad {\rm with} \quad
            \epsilon = \frac{E}{E_T} \,,\,\,
             E_T = \frac{\hbar v_F}{2 L_T} \,.    
\end{equation}
This result was also derived in \cite{Schom99} 
in the context of a chaotic billiard 
coupled to a superconducting lead of small width $w$.
It remains valid for billiards with boundary roughness as long as
the time scale $t_{\rm erg}$ on which the classical motion in the
billiard becomes ergodic is much smaller than the mean escape time
$\tau_{\rm esc}$. 

\begin{figure}
\begin{minipage}{16.6cm} 
  \centerline{\large Fig. 4}
  \vspace{1.0cm}
  \centerline{\psfig{figure=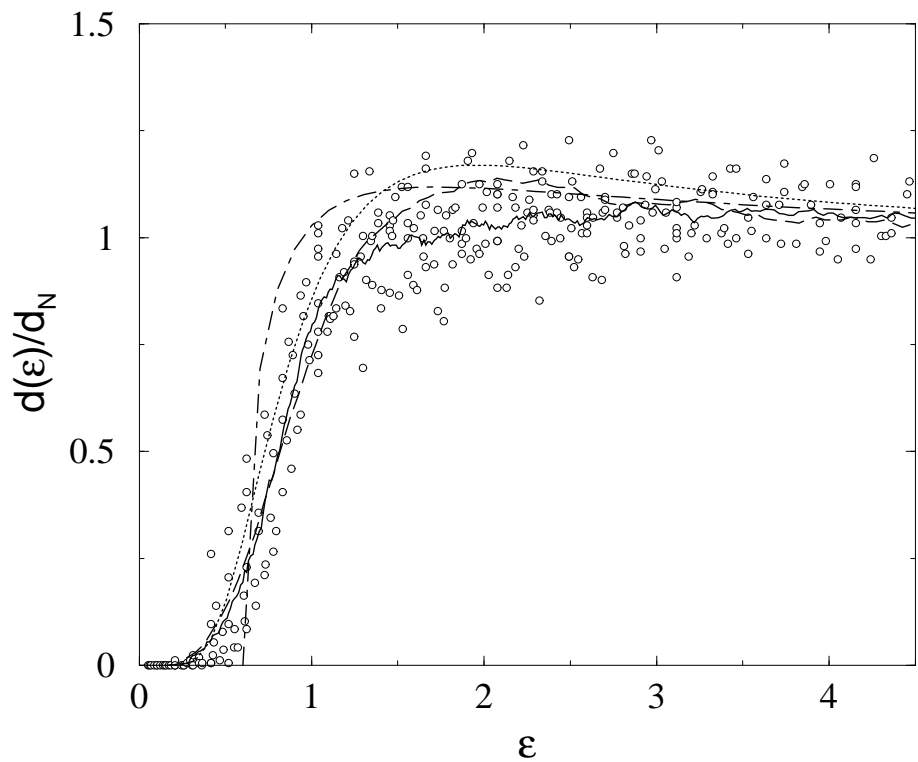,width=12.0cm}}
  \caption[Short title]{Average density of states for a rough square
                        billiard. Data points are quantum mechanical
                        energy eigenvalues for a 
                        billiard of side length $a=75$ and channel widths
                        $w \le 40$. The solid line is a 20-point
                        average of the numerical data. Dashed curve: 
                        semiclassical calculation based on
                        Eq. (\ref{eq:dosgen}). Dotted curve:
                        Analytical expression Eq. (30).
                        Dot-dashed line: Random matrix result of
                        Ref. [3].
                        }
   \label{fig4}
\end{minipage}
\end{figure}

The semiclassical prediction (\ref{eq:dosforrough}) is compared with
numerical quantum mechanical results in Fig. \ref{fig4}. A 20-point
average was taken over the numerical results for different channel
widths $w$ (solid curve). It is in very good agreement
with the analytical formula (\ref{eq:dosforrough}) (dotted curve)
as well as an evaluation of Eq. (\ref{eq:dosrough}) with numerically
calculated  length distribution functions $P(L)$ (dashed curve). The Thouless
energy is inversely proportional to twice the escape time $\tau_{\rm
esc}$ of an electron trajectory since for a complete transversal of
a path hitting the SN-boundary the electron part and the retracing 
as a hole trajectory have to be added for one full cycle. 

The exponential suppression of the spectral density for the rough
billiard in contrast to the linear rise in energy for the square
billiard without roughness has its origin in the asymptotic behaviour
of the length distribution function $P(L)$. The density of low-lying
quasiparticle excitations is determined by very long orbits in order
to fulfill the delta function condition in Eq. (\ref{eq:dosrough})
in the limit $E\to 0$. As a consequence of the exponential tail
of the length distribution (\ref{eq:lengthdistrough}) the density of
states is exponentially suppressed above the Fermi energy. In contrast
the algebraic tail of $P(L)$ in the square billiard without roughness
leads only to a linear suppression in energy.

In a different approach the Hamiltonian of an (isolated) chaotic
mesoscopic billiard was modelled by a GOE ensemble from random
matrix theory and its coupling to the superconducting leads was then
taken into account by means of a coupling matrix \cite{Mel96,Mel97}.
There it was shown that the quasiparticle density of states vanishes
{\it exactly\/} below an energy of approximately $E_c<0.6E_T$ (see
dashed-dotted line in Fig. 4). 
As discussed at the end of Sec. II the difference between the two
results can be traced back to the diagonal approximation employed in the
semiclassical theory. Despite this approximation
the quantum mechanical results of Fig. \ref{fig4}
are in good agreement with the semiclassical prediction of
an exponential suppression of the density of states. This is due to
the fact that the random matrix theory of \cite{Mel96} predicts a gap
in the excitation spectrum in the limit of an infinity number of
channels $N\to\infty$ (since it is essentially an expansion in the
parameter $1/N$). For a finite number of channels the behaviour of
the spectral density around $E_c$ is expected to be smoothed out \cite{Frah}.
\bigskip

\leftline{\bf 3.3 Effect of a magnetic field in rough billiards}
\bigskip

In this subsection we consider the effect of a magnetic field
on the density of quasiparticle states. The magnetic field $B$ is
uniform and points in the direction perpendicular to the billiard area.
The superconducting lead itself is not penetrated by the flux.
We study the perturbative regime of small magnetic
fields where the cyclotron radius
is much larger than the length scale $a$
defined by the size of the mesoscopic billiard itself. The
trajectories are then unchanged and the only effect of
the magnetic field is an additional phase acquired by each trajectory
which is 
proportional to the directed flux enclosed by the trajectory \cite{Ric96}.
Formula (\ref{eq:dosgen}) for the average density of states is then
modified to include the flux dependent phase for each trajectory.
The result is
\begin{equation}\label{eq:dosfluxinc}
            d(E) = \frac{d_N}{A} \int \limits_{0}^{W} dy
               \int\limits_{-1}^{1} d(\sin\alpha)
            \int\limits_{0}^{L(y,\alpha)} ds \sum_n
              \delta\left(\frac{E
            L(y,\alpha)}{\hbar v_F} -(n+\frac{1}{2})\pi
              + 2\pi\frac{\Phi(y,\alpha)}{\Phi_0}\right) \,,
\end{equation}
where $\Phi_0=ch/e$ is the flux quantum. Electron trajectories which
traverse the billiard in opposite directions between two hits with the
superconducting part of the boundaries acquire flux of same magnitude
but opposite sign. This corresponds to a splitting of energy levels linear in $B$
which are degenerate in the absence of a magnetic field \cite{Bru98}.
For the generic case of rough or chaotic billiards a statistical
description can be used in the same way as was done in the
previous section for the magnetic field free case. In addition to
the length distribution $P(L)$, which remains unchanged,
the distribution $P_L(\Theta)$ of the directed area $\Theta$ enclosed
by the trajectories of given length
$L$ must be specified. For long trajectories it is given by a Gaussian
of the form \cite{Bar93,Ric96}
\begin{equation}
        P_L(\Theta) = \frac{1}{\sqrt{2\pi L\sigma_L}}
       \exp\left(-\frac{\Theta^2}{2 L\sigma_L}\right) \,,
\end{equation}
where $\Theta$ is the directed area enclosed by the orbit.
A heuristic argument for this distribution goes like follows:
for a long orbit the length is on average proportional to the number
of bounces with the rough billiard walls. At each bounce the
trajectory is randomised and therefore the area swept between two
successive bounces can be viewed as a random variable with zero mean
value. Trajectory segments are independent of each other.
The total area $\Theta$ accumulated by a trajectory is therefore
a sum of independent random variables and its distribution is a Gaussian.
It follows that the area distribution integrated over length
has an exponential form:
 $P(\Theta) = (2{\bar\Theta})^{-1}\exp(-\Theta/{\bar\Theta})$
with ${\bar\Theta}=\sqrt{L_T\sigma_L/2}$. 

\begin{figure}
\begin{minipage}{16.6cm} 
  \centerline{\large Fig. 5}
  \vspace{1.0cm}
  \centerline{\psfig{figure=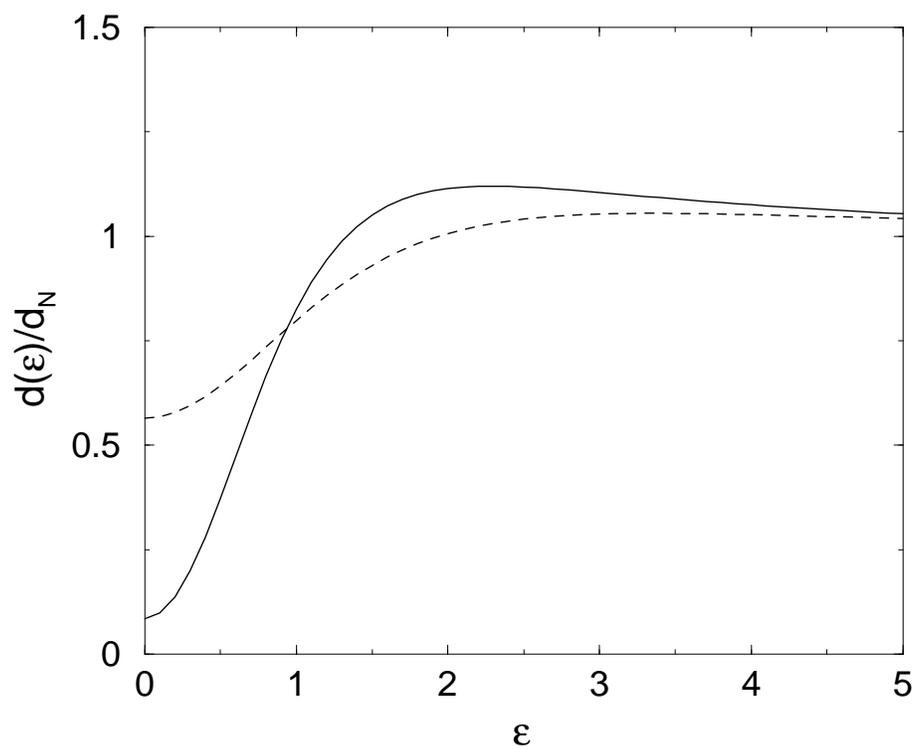,width=12.0cm}}
  \caption[Short title]{Average density of states $d(\epsilon)/d_N$
                       for the rough square billiard of unit area
                       in the presence of a flux according to
                       the semiclassical formula
                       (\ref{eq:dosfluxend}). Solid line:
                       $\phi=0.5$. Dashed line: $\phi=1.0$.}
   \label{fig5}
\end{minipage}
\end{figure}

Using the above area distribution and an ergodic distribution
of initial conditions along the superconducting channel boundary the density
of states can be rewritten as 
\begin{eqnarray}\label{eq:doswithflux}
       d(E,B) &=& \frac{d_N W}{A} \int_{0}^{\infty} dL\, P(L)\,L\,
              \int_{-\infty}^{\infty} d\Theta P_L(\Theta)
                \nonumber\\
            & &
            \times \sum_n \delta\left(\frac{E
              L}{\hbar v_F} -(n+\frac{1}{2})\pi +
               \frac{2\pi B\,\Theta}{\Phi_0}\right)  \nonumber
\end{eqnarray} 
Transforming the sum over delta functions by use of the Poisson formula
and performing the integration over areas and lengths successively one
finally gets the following expression for the average spectral density
of quasiparticle excitations as a function of energy and magnetic flux:
\begin{equation}\label{eq:dosfluxend}
       d(\epsilon,\phi) = d_N + 2 d_N \sum_{k=1}^{\infty} (-1)^k
             \frac{(1+k^2\phi^2)^2 - (k\epsilon)^2}{
                   [(1+k^2\phi^2)^2 + (k\epsilon)^2]^2} \,\,,\quad\quad
      \phi = \frac{4\pi B}{\Phi_0}\sqrt{2\sigma_L L_T} \,.
\end{equation}
The dimensionless quantity $\phi$ is the effective flux through the
billiard measured in units of the flux quantum $\Phi_0=ch/e$. This is
evident when it is written as $\phi=4\pi B {\bar\Theta}/\Phi_0$.
${\bar\Theta}$ has the meaning of an effective area. The dimensionless 
flux can also
be expressed as $\phi = \Phi_{\rm tot}/\Phi_{\rm cr}$, where
$\Phi_{\rm tot}=BA$ is the total flux through the billiard and
$\Phi_{\rm cr} = A/(4\pi {\bar\Theta}) \Phi_0$. As will be discussed
in the following in the regime $\Phi_{\rm tot}<\Phi_{\rm cr}$ ($\phi<1$)
the exponential suppression of the DOS near the Fermi energy persists
while for larger fluxes the DOS converges towards a constant value.
Fig. \ref{fig5} shows the quasiparticle excitation spectrum
as a function of energy for two different values of
the flux parameter $\phi$.
With growing flux the average density of states acquires the value $d_N$ 
of the isolated billiard also for values $E<E_T$.

At the Fermi energy $\epsilon=0$ Eq. (\ref{eq:dosfluxend}) can be
summed and gives
\begin{equation}\label{eq:dosfluxzero}
     d(\epsilon=0,\phi) = \frac{d_N}{2} \frac{\pi}{\phi}
                         \left[\sinh\frac{\pi}{\phi}\right]^{-1}
             \left[\frac{\pi}{\phi}\coth\frac{\pi}{\phi}
                 + 1\right]
              \,.
\end{equation}
Eq. (\ref{eq:dosfluxzero}) should not be understood in the sense 
of an expression for the density of states {\it exactly\/} at the Fermi 
level. Properly interpreted 
it is {\it proportional\/} to the quasiparticle density of states
as a function of the magnetic flux when the density of states is
averaged over an energy window $0 \le E \le \delta$ between the Fermi energy
and the mean level spacing $\delta$ of the isolated billiard. 
(Note that the averaging is over an energy scale much smaller than $E_T$.)
It has been shown that even at fluxes $\phi\gg 1$
a minigap of order of the mean level spacing is present in the
quasiparticle density of states at the Fermi
level \cite{Fra96,Altl96}. 
Therefore the quasiparticle DOS exactly
at the Fermi level is always zero. According
to Ref. \cite{Fra96} for a billiard whose isolated dynamics is
described by a GUE matrix ensemble (which corresponds to $\phi>1$)
the DOS is given
by $d_{\rm GUE}(E)/d_N = 1-\sin(2\pi d_N E)/(2\pi d_N E)$. 
When $d_{\rm GUE}(E)/d_N$ is averaged over the interval $0<E<\delta$ it gives the
value $g=0.77$. Using this asymptotic value for large fluxes we therefore
expect the DOS $d_\delta$ averaged over the window $0 \le E \le \delta$
to be given by
\begin{equation}\label{eq:dosfluxzero2}
       d_\delta(0,\phi) =  g d(0,\phi) \,.
\end{equation}
The factor $g$ has the effect that $d_\delta(0,\phi)$ 
saturates at a value smaller than $1$ in large magnetic fields, which
is the consequence of the existence of the minigap.

\begin{figure}
\begin{minipage}{16.6cm} 
  \centerline{\large Fig. 6}
  \vspace{1.0cm}
  \centerline{\psfig{figure=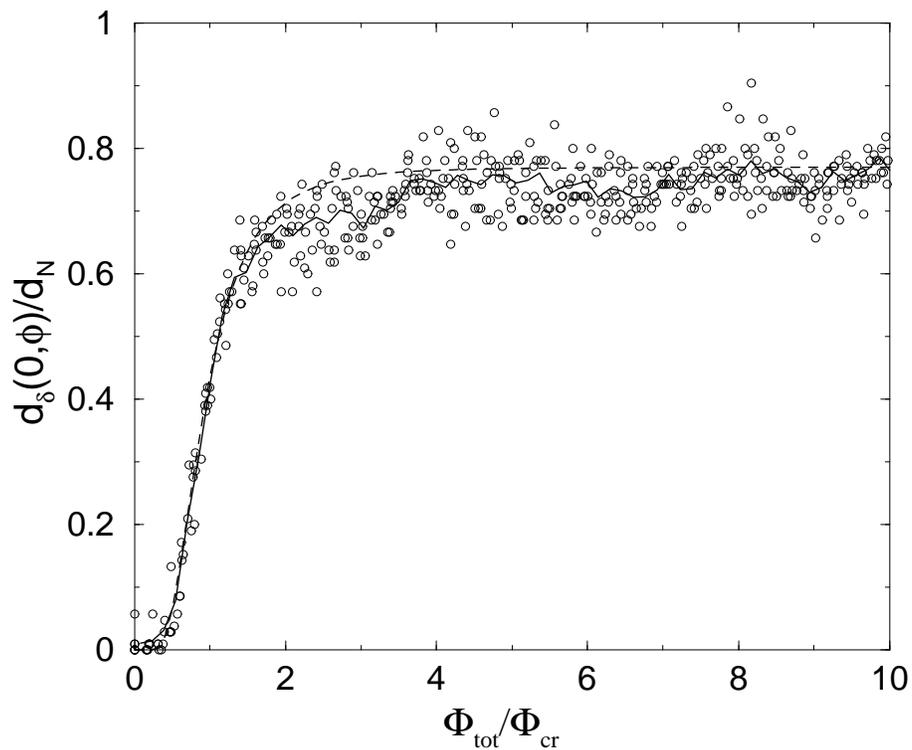,width=12.0cm}}
  \caption[Short title]{Spectral density $d_\delta(0,\phi)/d_N$ of 
                        quasiparticle excitations as a function of
                        the flux $\phi=\Phi_{\rm tot}/\Phi_{\rm cr}$
                        for the rough square billiard. The
                        density of states is averaged over the small
                        energy interval $0\le E\le\delta$, where
                        $\delta$ is the mean level spacing of the
                        isolated billiard. Data points: quantum mechanical
                        calculation width 10 different channel width $w$.
                        Solid line: Average over quantum mechanical
                        data points. Dashed line: semiclassical
                        theory, Eq. (\ref{eq:dosfluxzero}) and
                        (\ref{eq:dosfluxzero2}).
      }
   \label{fig6}
\end{minipage}
\end{figure}

Fig. \ref{fig6} compares the theory of
Eq. (\ref{eq:dosfluxzero2}) (dashed
line) with a numerical quantum mechanical
calculation (solid line). It shows that $d_\delta(0,\phi)$ is
exponentially suppressed on the scale $\phi<1$. The quantum mechanical
results were obtained for a rough billiard (sides $a=1$ and
$b=1.5$) with 10 different values of the width between $w=0.2$ and
$w=0.38$. In Ref. \cite{Bar93} numerical evidence was put forward 
that the effective area scales like ${\bar\Theta}=\alpha_0
A^{5/4}w^{-1/2}$ with the parameters of the billiard ($\alpha_0$ is
a numerical parameter). The flux scale $\Phi_{\rm cr}$ entering into
the magnetic field dependence of the DOS then scales as
$\Phi_{\rm cr}=(4\pi\alpha_0)^{-1} w^{1/2}A^{-1/4} \Phi_0$. The quantum
mechanical calculations for different channel widths $w$ and fixed
billiard area $A$ confirm this scaling property. Determining
$\alpha_0=0.1$ from the numerical data the flux scale $\Phi_{\rm cr}$
is fixed for each value of $w$. The quantum mechanical data points
plotted as a function of the flux $\Phi_{\rm tot}/\Phi_{\rm cr}$
and its average (solid line) are in excellent agreement with the
semiclassical theory of Eq. (\ref{eq:dosfluxzero2}). 

The difference between Eq. (\ref{eq:dosfluxzero}) which would lead to 
a finite density of states at $\epsilon=0$ at nonzero flux and the
exact result of Ref. \cite{Fra96} which predicts the minigap 
demonstrates the importance of nondiagonal contributions in the
trace formula (\ref{eq:avdens}) for energies $E<\delta$.
As it is known the diagonal approximation, which we employed, 
breaks down on the scale of the mean level spacing \cite{Ber85,Bog96}
and therefore semiclassics within the diagonal approximation does
not lead to the correct result on the scale of the mean level spacing
itself. When the density of states is averaged over a scale of the
mean level spacing $\delta$ or larger Eq. (\ref{eq:dosfluxzero2})
describes the behaviour of the spectral density correctly as discussed
above.

Finally we would like to emphasize that the semiclassical theory
of the proximity effect on the density of states in rough or chaotic
Andreev billiards allows for a very intuitive picture of its destruction
in the presence of a magnetic flux. The suppression of the density
of states below $E_T$ is a consequence of the coupling between
electron and hole like quasi-particles at the SN-interface. Classically
the return probability of trajectories to the SN-interface is given
by $P(L)$. Without magnetic field a path and its reverse interfere
constructively in Eq. (\ref{eq:dosrough}) for the density of
states. In the presence of a magnetic flux the 
phase difference between a path and its reverse leads to destructive
interference. Effectively the quantum
mechanical return probability is reduced compared to its classical
value $P(L)$. Stated differently the coupling of the mesoscopic
billiard to the superconducting lead (the possibility of electron hole
conversion) is reduced in the presence of a magnetic flux and thus
the proximity effect is suppressed. With growing magnetic flux the
billiard will effectively look like an isolated mesoscopic conductor
and its average density of states will therefore acquire the value $d_N$.
The effect is reminiscent of the weak localisation correction to the
classical reflection coefficient \cite{Bar93} and of enhanced orbital
magnetism \cite{Ric96} in ballistic semiconductor microstructures.
\bigskip

\leftline{\bf 4 Summary}
\bigskip

We studied the quasiparticle excitation spectrum of a mesoscopic
conductor modelled by a billiard in proximity to a superconducting
lead. The expression for the density of states was derived from the
semiclassical scattering matrix formulation. It was shown to be equivalent
to the result derived previously from the Eilenberger equation for the
quasiparticle Greens function when the diagonal
approximation is applied to traces of powers of the scattering matrix.
Applications to a square billiard as an
example of a classical integrable system and a square billiard with
surface roughness as an ergodic system were considered. For the square
billiard without roughness it was shown that the deviation of the
length distribution function of trajectories hitting the
superconducting part of the boundary at small lengths is responsible
for the saturation of the average density of states at large energies.
For the rough billiard semiclassics predicts an
exponential suppression of the density of states at energies smaller
than the Thouless energy. The difference to the random matrix result
of an energy gap of order of the Thouless energy can be traced back 
to the diagonal approximation employed in the semiclassical theory.
The semiclassical predictions are in 
good agreement with quantum mechanical calculations for the integrable
as well as for the ergodic billiard. Finally we derived 
expressions for the effect of a weak magnetic field on the density of
states in the rough billiard. A magnetic field destroys the proximity
effect on the density of states. We interpreted this fact as a phase
phenomenon involving identical but reversed paths
which hit the superconducting part of the billiard boundary.
Due to destructive interference in the presence of a magnetic flux
the quantum mechanical return probability of trajectories to the
superconducting lead is smaller than its classical value 
effectively reducing the coupling between the billiard and the
superconducting lead. 
\bigskip

\small{ We enjoyed discussions with A. Altland, C. Beenakker, K. Frahm, B. Mehlig
and H. Schomerus and thank M. Sieber for a critical reading of the manuscript.
WI would especially like to thank F. Mota-Furtado and P. F. O'Mahony for
encouraging him to work on Andreev billiards and for discussions on the
subject in the early stage of this work.}

\vfill\eject
\vfill\eject
\vfill\eject

\vfill\eject
\vfill\eject

\vfill\eject

\vfill\eject

\end{document}